\documentclass[prd,aps,twocolumn,10pt,showpacs,nofootinbib]{revtex4}

\usepackage{epsfig}
\usepackage{enumitem}

\newcommand{\be}{\begin{equation}}
\newcommand{\ee}{\end{equation}}
\newcommand{\bea}{\begin{eqnarray}}
\newcommand{\eea}{\end{eqnarray}}
\newcommand{\bc}{}

\begin{document}

\title{FRW cosmology in Milgrom's bimetric theory of gravity}
\author{Timothy~Clifton$^{1}$\footnote{tclifton@astro.ox.ac.uk} and Thomas~G.~Zlosnik$^{2}$\footnote{tzlosnik@perimeterinstitute.ca}}
\affiliation{
$^1$Department of Astrophysics, University of Oxford, UK.\\
$^2$Perimeter Institute of Theoretical Physics, Waterloo, Canada.\\
}
\pacs{98.80.jk, 04.20.Jb, 04.50.Kd}
\begin{abstract}
We consider spatially homogeneous and isotropic
Friedmann-Robertson-Walker (FRW) solutions of Milgrom's recently proposed
class of bimetric theories of gravity.  These theories have two
different regimes, corresponding to high and low acceleration.  We
find simple power-law matter dominated solutions in both, as well as
solutions with spatial curvature, and exponentially expanding
solutions.  In the high acceleration limit these solutions behave like
the FRW solutions of General Relativity, with a cosmological constant
term that is of the correct order of magnitude to explain the observed accelerating expansion of the Universe.
We find that solutions that remain in the high acceleration regime for their entire history, however,
require non-baryonic dark matter fields, or extra interaction terms in
their gravitational Lagrangian, in order to be observationally viable.  The low
acceleration regime also provides some scope to account for this deficit,
with solutions that differ considerably from their general
relativistic counterparts.
\end{abstract}

\date{May 21, 2010}
\maketitle

\section{Introduction}

A natural way to extend General Relativity is via the introduction of a second
dynamical rank 2 metric tensor field, $\hat{g}_{ab}$.  The so called `bimetric' theories that result
have a rich phenomenology (see, for example, \cite{p1,p2,p3,p4}), and a long
history \cite{h1,h2,h3}.  They represent an application to gravity of the
Yang-Mills approach to gauge theories, and allow for new and
interesting behaviour.

A new class of these theories has recently been introduced by Milgrom
\cite{Milgrom1, Milgrom2}, in which the interaction term between the
two metric fields is constructed from a tensor defined as
the difference of the connections associated with each of
them.  The stated purpose of this approach is to produce a weak-field limit of the form
\be
\label{mondp}
\nabla^2 \Psi
-\nabla\cdot\left[f\left(l |\nabla\Psi_N|\right)\nabla\Psi_N\right]
= 4\pi G \rho,
\ee
where $\Psi_N$ obeys $\nabla^2
\Psi_N=4\pi G \rho$, and where time-like test particles obey a force law $\ddot{x} \sim -\nabla
\Psi$. Here $\Psi$ is a gravitational potential, $f(x)$ is a function to be
specified, $\rho$ is the energy density and $l$ is a constant with
units of length.  The existence of gravitational fields of
this type would be of considerable interest for astrophysics
\cite{new,Milgrom83}, and the goal of producing a viable relativistic
formulation for theories of this kind has been pursued for some time
\cite{rel1,rel2,rel3,rel4,rel5,rel6,rel7,rel8,rel9}.

%

Let us now try to understand why a relativistic formulation of (\ref{mondp}) is so
important.  If we associate an energy-momentum tensor, $T_{ab}$, with
the right-hand side of (\ref{mondp}), then
\be
\rho = T_{ab}n^{a}n^{b},
\ee
where $n^{a}$ is the unit vector normal to surfaces of constant $t$.
Energy-momentum conservation then tells us that test particles follow
geodesics of the metric with respect to which $T_{ab}$ is covariantly
conserved. For a perturbed metric of the form
\begin{eqnarray}
\label{conform}
ds^{2} &=& -(1+2\Psi)dt^{2} + (1-2\Phi)\delta_{ij}dx^{i}dx^{j}
\end{eqnarray}
this corresponds to $\ddot{x} \sim -\nabla\Psi$, as required.
In order to gain a set of field equations that are invariant with
respect to coordinate transformations, and that reduce to
(\ref{mondp}) for the metric (\ref{conform}), we must now construct a
tensor quantity that can be associated with the left-hand side of (\ref{mondp}).
We immediately encounter a problem though:
The quantity $|\nabla\Psi|$ cannot be formed from any local curvature
invariants of (\ref{conform}) alone. One may alternatively try and recover (\ref{mondp}) 
from 
\emph{nonlocal} functions of the space-time curvature \cite{sous1}. However,
such attempts have prompted a no-go theorem \cite{sous2} regarding the phenomenological
viability of all theories that are constructed from the space-time
geometry and matter fields only.
New structure is therefore required, beyond a single space-time metric.

In the absence of a relativistic theory that is well motivated by
other considerations, and that reduces to (\ref{mondp}) in the 
appropriate limit, it is not clear what this additional
structure should be.  We are then left without any way of knowing when
(\ref{mondp}) should be considered valid, or when we should prefer,
for example, a Newtonian description. We can, however, use observations of
relativistic effects beyond Newtonian order to gain insight.  In this regard, cosmological solutions provide
us with a very useful probe as they express the full non-linearity of
the relativistic field equations.  Motivated by this, as well as the
desire to understand further the cosmological consequences of bimetric
theories of gravity in general, we consider here the FRW solutions of
Milgrom's class of bimetric theories.

Of course, for any theory of gravity to be considered viable it is now
the case that it should be able to reproduce all of the major probes of
observational cosmology.  These include the
anisotropies in the cosmic microwave background, the matter power
spectrum on large scales, Hubble diagrams that extend out to
$z\sim 1$, as well as the ratios of light elements from primordial nucleosynthesis.  All
of these require a detailed knowledge of FRW
cosmology (as well as of cosmological perturbation theory, in the
case of the first two).  It is not our goal here to perform an
exhaustive study of all of these areas, but rather to make a solid
first step in understanding the underlying FRW solutions, and what
they imply.
%

The article proceeds as follows.  In Section \ref{field} we describe the theory and its field
equations. The general form of the theory
allows considerable flexibility, and so in Section \ref{spec} we discuss and
motivate the specific versions of the theory that we will consider. In
Section \ref{FRWsec} we find the field equations in space-times with
FRW symmetries, and discuss their solutions in Section
\ref{solutions}. In Section \ref{viao} we discuss the degree to which
current probes of cosmology can constrain these models, and in Section
\ref{disc} we conclude.

\section{Field Equations}
\label{field}

The theory we are considering has two metrics, $g_{ab}$ and
$\hat{g}_{ab}$. The action for the gravitational part of Milgrom's theory
can be written in terms of these two tensor fields as
\be
I = I_{g}+I_{\hat{g}}+I_{int},
\ee
where
\begin{eqnarray}
\label{I1}
I_{g} &=& \beta \int  \sqrt{-g} R d^{4}x \\
\label{I2}
I_{\hat{g}} &=& \alpha \int \sqrt{-\hat{g}} \hat{R} d^{4}x
\end{eqnarray}
are the Einstein-Hilbert terms for $g_{ab}$ and $\hat{g}_{ab}$, and
$I_{int}$ is an interaction term between them.  In the equations
above $R$ is the Ricci scalar constructed from $g_{ab}$, and $\hat{R}$ is
the Ricci scalar constructed from $\hat{g}_{ab}$.

In Milgrom's theory the interaction term, $I_{int}$, is specified as a
function of scalars formed from the rank-3 tensor
\be
C^a_{\phantom{a} b c} = \Gamma^a_{\phantom{a} b
 c}-\hat{\Gamma}^a_{\phantom{a} b c}, 
\ee
where $\Gamma^a_{\phantom{a} b c}$ and $\hat{\Gamma}^a_{\phantom{a} b
 c}$ are the metric connections of $g_{ab}$ and $\hat{g}_{ab}$,
respectively, such that
\bea
g_{ab;c} &=& 0\\
\hat{g}_{ab:c} &=& 0,
\eea
where $;$ and $:$ indicate covariant derivatives constructed from
$\Gamma^a_{\phantom{a} b c}$ and $\hat{\Gamma}^a_{\phantom{a} b c}$.
The general form of scalars that are quadratic in $C^a_{\phantom{a} b
 c}$ can then be written as
\begin{eqnarray}
Q &=&  l^2 Q_{ad}^{\phantom{ad}bcef}C^{a}_{\phantom{a}bc}C^{d}_{\phantom{d}ef},
\end{eqnarray}
where $Q_{ad}^{\phantom{ad}bcef}$ is built from $g_{ab}$ and
$\hat{g}_{ab}$, and $l$ is a constant with dimensions of length.  In
terms of these quantities we can then write\footnote{Note that in terms
of the formalism used by Milgrom $\sigma=f(\kappa)/\kappa$, and $Q=l^2
\Xi$.}
\be
\label{interaction}
I_{int} = \frac{2}{l^2} \int \sqrt{-g} \sigma \mathcal{M}(Q) d^{4}x,
\ee
where $\mathcal{M}(Q)$ and $\sigma=\sigma(\kappa)$ are functions to be
specified, and $\kappa=(g/\hat{g})^{1/4}$.

In Appendix A we find the field equations for $g_{ab}$ can be written as
\begin{eqnarray}
\label{field1}
\beta G_{ab} +{\cal S}_{ab}&=& 8 \pi G T_{ab} 
\end{eqnarray}
where
\begin{eqnarray*}
{\cal S}_{ab}&=&2 \left[\sigma
 \mathcal{M}_{Q}\left(J_{(ab)}^{\phantom{(ab)}c}+J_{(a \phantom{c}b)}^{\phantom{(a}c}
	    -J^{c}_{\phantom{e}ab}
\right)\right]_{;c}
\\
\nonumber && +
2 \sigma \mathcal{M}_{Q}\frac{\delta
 Q_{gd}^{\phantom{gd}hcef}}{\delta
 g^{ab}}C^{g}_{\phantom{a}hc}C^{d}_{\phantom{d}ef}
-\frac{\sigma(1+n)}{l^2}\mathcal{M}
 g_{ab},
\end{eqnarray*}
and $G_{ab}$ and $T_{ab}$ are the Einstein tensor and the
energy-momentum tensor, defined with respect to
$g_{ab}$ in the usual way.  We have also used the notation
$\mathcal{M}_{Q}\equiv d\mathcal{M}/dQ$ and $J_{d}^{\phantom{d}ef}
\equiv Q_{ad}^{\phantom{ad}bcef} C^{a}_{\phantom{a}bc}$, and taken
$\sigma= \kappa^{2n}$.

Similarly, the field equations for $\hat{g}_{ab}$ are found to be
\begin{eqnarray}
\label{field2}
\alpha\hat{G}_{ab}+\hat{{\cal S}}_{ab} &=& 8 \pi G \hat{T}_{ab}
\end{eqnarray}
where
\begin{eqnarray*}
\nonumber
\hat{\cal S}_{ab}&=&-2 \left[\sigma^{\frac{n+1}{n}}
 \mathcal{M}_{Q}\left(2
 J_{(a}^{\phantom{a}(ef)}\hat{g}
 \vphantom{J}_{b)f}^{\phantom{a)e}}-J_{d}^{\phantom{d}cf}(\hat{g}^{-1})^{de}\hat{g}_{ca}\hat{g}_{fb}\right)\right]_{:e} \\
\nonumber &&+2\sigma^{\frac{n+1}{n}} \mathcal{M}_{Q}\frac{\delta
 Q_{yd}^{\phantom{yd}zcef}}{\delta
 \hat{g}^{ab}}C^{y}_{\phantom{a}zc}C^{d}_{\phantom{d}ef}+\frac{\sigma^{\frac{n+1}{n}} n}{l^2}\mathcal{M} \hat{g}_{ab},
\end{eqnarray*}
and where $\hat{G}_{ab}$ and $\hat{T}_{ab}$ are the Einstein and
energy-momentum tensors defined with respect to $\hat{g}_{ab}$.  Here,
and throughout, we write the inverse of
$\hat{g}_{ab}$ as $(\hat{g}^{-1})^{ab}$ (defined such that $(\hat{g}^{-1})^{ac}\hat{g}_{bc} =
\delta^a_{\phantom{a} b}$).  Any raising or lowering of indices is
otherwise always done with $g_{ab}$.

It should be noted that in (\ref{field1}) and (\ref{field2}) above we
have taken the matter fields described by $T_{ab}$ and $\hat{T}_{ab}$
to be coupled to $g_{ab}$ and $\hat{g}_{ab}$, respectively\footnote{Although
one could also concievably couple matter to a combination of these
metrics, we prefer to restrict ourselves here to the case considered
by Milgrom in \cite{Milgrom1, Milgrom2}.}.  We presume that we (as
observers) are made from matter coupled to only one of these metrics,
which we take to be the former without loss of generality.  We further
presume that there is no interaction between $T_{ab}$ and
$\hat{T}_{ab}$, so that all observations we make will be of the other
matter fields coupled to $g_{ab}$.  Cosmological probes of the
expanding Universe then give us direct information about the geometry
of $g_{ab}$ only.

In the equations derived above we have not explicitly included any
cosmological constant terms.  However, from (\ref{field1}) and
(\ref{field2}) it is clear that $\Lambda$ is dynamically
equivalent to a perfect fluid with $p=-\rho$.  We therefore account for
any possible cosmological constants by allowing for them in $T_{ab}$
and $\hat{T}_{ab}$.

\section{Specification of the Theory}
\label{spec}

In Section \ref{field} we presented the field equations for the theory
derived from the action specified by (\ref{I1}), (\ref{I2}) and
(\ref{interaction}).  These equations provide constraints on, and specify
the evolution of, the two dynamical rank two tensor fields $g_{ab}$
and $\hat{g}_{ab}$, but they also contain considerable freedom:  The
tensor $Q_{ab}^{\phantom{ab}cdef}$ and the function $\mathcal{M}(Q)$
have yet to be specified.  In order to make progress in
understanding the cosmological solutions of these theories, we will
therefore now restrict ourselves to specific cases.

Firstly, we will only consider
$Q_{ab}^{\phantom{ab}(cd)(ef)}=Q_{ba}^{\phantom{ab}(ef)(cd)}$ that are
formed from the metric tensor $g_{ab}$.  In this case the most
general $Q_{ab}^{\phantom{ab}cdef}$ that can be constructed is
\bea
\nonumber
Q_{ab}^{\phantom{ab}cdef} &=& c_1 \delta^f_{\phantom{f} a}
 \delta^d_{\phantom{d} b} g^{c e} + c_2  \delta^f_{\phantom{f} b}
     \delta^e_{\phantom{e} a} g^{cd} + c_3 g^{cd} g^{ef} g_{ab} \\&&+c_4
	 \delta^c_{\phantom{c} a}  \delta^e_{\phantom{e} b} g^{df}
	     +c_5 g_{ab} g^{ce} g^{d f},
\label{Qabcdef}
\eea
where $c_1$, $c_2$, $c_3$, $c_4$ and $c_5$ are constants, and indices
are assumed to be symmetrized appropriately.  In this
notation, the `concrete simple theory' of Milgrom is specified by
$c_1=1$, $c_2=-1$ and $c_3=c_4=c_5=0$.  The form of
$Q_{ab}^{\phantom{ab}cdef}$ given in (\ref{Qabcdef}) can then be seen
to correspond to a generalization of this theory.

Secondly, we will consider the function $\mathcal{M}(Q)$ to have the
form specified by Milgrom in the low and high acceleration limits of
the theory.  That is, when $\vert Q \vert \gg 1$ we take
\be
\label{newton}
\mathcal{M}_Q \simeq 0,
\ee
and when $\vert Q \vert \ll 1$ we take
\be
\label{mond}
\mathcal{M}_Q \simeq \vert Q \vert^{-1/4}.
\ee
The first of these corresponds to Newtonian gravitation in the
non-relativistic limit (when $c\rightarrow \infty$), and the second
corresponds to modified gravitational dynamics in the low acceleration
regime of the non-relativistic limit (when $c\rightarrow \infty$ and
$l \rightarrow 0$).  In particular, when $\alpha+\beta=0$ and $\hat{T}_{ab} =0$ one
recovers (\ref{mondp}) in the weak field limit, with $f=\mathcal{M}_Q$
\cite{Milgrom1}.  The weak field limit with more general $\alpha$ and
$\beta$, and with $\hat{T}_{ab}$, has been explored in \cite{Milgrom2}.

Here we will not be concerned, for the most part, with the transitional
behaviour between these regimes. It is not clearly specified by the
weak field limit of the theory, and is presumed to be highly sensitive
to the particular form of $\mathcal{M}(Q)$ that is chosen.

\section{FRW Cosmology}
\label{FRWsec}

We will now consider the cosmological solutions of the theory discussed above.
Imposing FRW symmetries on $g_{ab}$ and $\hat{g}_{ab}$ leads to the line-elements
\be
\label{ds}
ds^2 = g_{ab}dx^{a}dx^{b} = -a(\tau)^2
d\tau^{2}+a(\tau)^{2}h_{ij}dx^{i}dx^{j},
\ee
and
\be
\label{ds2}
d\hat{s}^2 = \hat{g}_{ab}dx^{a}dx^{b} = -X(\tau)^{2}d\tau^{2}+Y(\tau)^{2}\hat{h}_{ij}dx^{i}dx^{j},
\ee
where $h_{ij}$ and $\hat{h}_{ij}$ are the metrics of
Euclidean 3-spaces of constant curvature, $k$ and $\hat{k}$, respectively.

The reason for introducing
the function $X(\tau)$ in the $\tau$-$\tau$ component of $\hat{g}_{ab}$ is that
we now only have one coordinate freedom of the form $\tau\rightarrow
f(\tau)$, but two $\tau$-$\tau$ components, in the two different metrics.  It is
therefore not possible to redefine $\tau$ to absorb $X(\tau)$ without
introducing a second function into the $\tau$-$\tau$ component of $g_{ab}$.
The most general expression of FRW geometry for both metrics must therefore
contain three functions of $\tau$: $a(\tau)$, $X(\tau)$ and $Y(\tau)$.

Using the geometry specified by (\ref{ds}) and (\ref{ds2}) we can now
calculate the quantities that appear in the field equations
(\ref{field1}) and (\ref{field2}).  Of particular interest is the
constraint equation, which contains only first derivatives of $a$,
$X$ and $Y$.  This equation is the analogue of the Friedmann equation
in general relativistic FRW cosmology, and is derived in Appendix B.
As an example, for Milgrom's `concrete simple theory' with
$\hat{k}=k$ (which we will motivate later) we find
\begin{widetext}
$$
\beta \left(\frac{\dot{a}^2}{a^2}+k\right) + \alpha \frac{Y^3}{a^2 X}\left(\frac{\dot{Y}^2}{Y^2}+k\frac{X^2}{Y^2}\right)
+\frac{a^2 \mathcal{M} \sigma}{3 l^2} +2
\sigma \mathcal{M}_Q
\left(\frac{\dot{Y}}{Y}-\frac{\dot{a}}{a} \right) \left[2
\frac{\dot{a}}{a}-\frac{\dot{X}}{X}+ \frac{\dot{Y}}{Y} -2 \frac{Y
 \dot{Y}}{X^2} \right] =\frac{8 \pi G
\rho a^2}{3}+ \frac{8 \pi G \hat{\rho} X Y^3}{3 a^2},
$$
where
$$
Q = \frac{3 l^2}{a^2} \Bigg[ 2 \frac{\dot{a}}{a} \frac{\dot{Y}}{Y} -2
 \frac{\dot{a}^2}{a^2} +2 \frac{Y \dot{Y}}{X^2} \frac{\dot{a}}{a}  
 + \frac{\dot{X}}{X} \frac{\dot{Y}}{Y} - \frac{Y\dot{Y}\dot{X}}{X^3}
 -\frac{\dot{Y}^2}{X^2}  -
 \frac{\dot{Y}^2}{Y^2}\Bigg],
\qquad \qquad \text{and} \qquad \qquad
\sigma =\left( \frac{a^4}{X Y^3}\right)^n \qquad \qquad \qquad
$$
\end{widetext}
and where $\rho$ and $\hat{\rho}$ are
the energy densities of the perfect fluids associated with the
energy-momentum tensors $T_{ab}$ and $\hat{T}_{ab}$, respectively.
Over-dots denote differentiation with respect to $\tau$.
More general expressions for the constraint equation, and for the other
quantities involved in the field equations, are given in Appendix C.

The field equations (\ref{field1}) and (\ref{field2}) also provide
second-order evolution equations for the three variables $a$, $X$ and
$Y$.  However, these expressions are very lengthy, and so we choose
not to reproduce them explicitly here.

As usual, the energy-momentum tensors obey conservation equations.  For
non-interacting fluids these equations read
\be
\label{con1}
\dot{\rho} + 3 \frac{\dot{a}}{a} (\rho+p) = 0
\ee
and
\be
\label{con2}
\dot{\hat{\rho}} + 3 \frac{\dot{Y}}{Y} (\hat{\rho}+ \hat{p}) = 0
\ee
where $p$ and $\hat{p}$ are the pressures of the two ideal
fluids\footnote{It is, in principle, possible to have an interaction between
the two fluids such that energy-momentum could be exchanged between
them.  We do not consider this possibility here, but rather treat the
two fluids as being non-interacting, as is usual in cosmology.}.  It
is convenient to treat these fluids as being polytropic, with constant equations
of state $w$ and $\hat{w}$ defined by $p = w \rho$ and 
$\hat{p} = \hat{w} \hat{\rho}$.  Equations (\ref{con1}) and
(\ref{con2}) then give $\rho \propto a^{-3(1+w)}$ and $\hat{\rho} \propto Y^{-3(1+\hat{w})}$.

\section{Solutions}
\label{solutions}

We will now look for solutions to the field equations in both the `high
acceleration' limit, when $\vert Q \vert\gg 1$, and the `low
acceleration' limit, when $\vert Q \vert \ll 1$.

\subsection{High acceleration limit}

For $\vert Q \vert \gg 1$ we have $\mathcal{M}_Q\simeq 0$, so the constraint and
evolution equations for $g_{ab}$ are given by
\bea
\label{F1}
\frac{\dot{a}^2}{a^4} &=& \frac{8 \pi G \rho}{3 \beta}-\frac{k}{a^2} -
\frac{\mathcal{M}_0 (1+n)\sigma}{3 \beta l^2}\\
\label{F2}
\frac{\ddot{a}}{a^3} - \frac{\dot{a}^2}{a^4} &=& - \frac{4 \pi G
 (\rho+3p)}{3 \beta} -
\frac{\mathcal{M}_0 (1+n)\sigma}{3 \beta l^2},
\eea
while the constraint and evolution equations for $\hat{g}_{ab}$ are
given by
\bea
\frac{\dot{Y}^2}{X^2Y^2} &=& \frac{8 \pi G \hat{\rho}}{3 \alpha}-\frac{\hat{k}}{Y^2} +
\frac{\mathcal{M}_0 n \sigma^{\frac{(1+n)}{n}}}{3 \alpha l^2}\\
\frac{\ddot{Y}}{X^2Y} - \frac{\dot{X}\dot{Y}}{X^3Y} &=& - \frac{4 \pi G
 (\hat{\rho}+3\hat{p})}{3 \alpha} +
\frac{\mathcal{M}_0 n\sigma^{\frac{(1+n)}{n}}}{3 \alpha l^2}
\eea
where
\be
\sigma=\left(\frac{\sqrt{1-\hat{k}
   r^2}}{\sqrt{1-k r^2}} \frac{a^4}{XY^3} \right)^n,
\ee
and $\mathcal{M}_0$ is the constant value of $\mathcal{M}(x)$ when
$x\gg 1$.  It is now clear that one or more of the following
conditions must be true:

\begin{enumerate}[label=$(\roman{*})$]
\item The value of $\mathcal{M}_0$ is $0$.
\item The value of $n$ is $0$ or $-1$.
\item The value of $\hat{k}$ is equal to that of $k$.
\end{enumerate}

If none of these conditions are met then the terms containing $\sigma$
have an $r$ dependence that cannot be accommodated by any of the
other terms.  The first two of these are conditions on the theory.  If
either, or both, of these conditions are met then there can exist FRW
solutions in the two metrics which have different spatial curvatures.
If neither $(i)$ nor $(ii)$ are satisfied, then the only FRW
solutions that exist are those in which the spatial curvatures of the
two FRW metrics are the same.

If condition $(i)$ is met then the Friedmann equations above can be
seen to be unchanged from their general relativistic form.  The two
scale factors $a$ and $Y$ are completely decoupled, and obey constraint and evolution equations that
are exactly the same as an FRW geometry in General Relativity.

If condition $(ii)$ is true (but $(i)$ is not) then either $a$ or $Y$
is driven by an additional term in its field equations that acts in exactly the same as a cosmological
constant.  For $n=0$ this term can be seen to appear in the Friedmann
equations for $a$, which are otherwise identical to their general
relativistic counterparts. The additional effective
cosmological constant is given by
\be
\Lambda_{eff} = - \frac{\mathcal{M}_0}{\beta l^2}.
\ee
If this is the case then $Y$ obeys field equations that are unchanged from a
general relativistic FRW solution.  If $n=-1$ then the situation is reversed,
with $Y$ being driven by an additional term, while the Friedmann
equations for $a$ are unchanged.

If the theory does not satisfy conditions $(i)$ or $(ii)$ then the
only FRW solutions that exist have $\hat{k}=k$.  The value of $\sigma$
is then independent of $r$, and $a$ obeys a set of Friedmann equations
with an additional effective
fluid whose energy density and pressure are given by
\be
p_{eff}=-\rho_{eff} = \frac{\mathcal{M}_0 (1+n)}{8 \pi G l^2}
\left(\frac{a^4}{XY^3} \right)^n
\ee
This effective fluid has an equation of state $w_{eff}=-1$, and
so we must have $\rho_{eff}=$constant, and therefore $a \propto X^{1/4}
Y^{3/4}$.  In this case the other set of Friedmann equations also has an additional
effective fluid, with
\be
\hat{p}_{eff}=-\hat{\rho}_{eff} = -\frac{\mathcal{M}_0 n}{8 \pi G l^2}
\left(\frac{a^4}{XY^3} \right)^{1+n}.
\ee
We therefore have that one of the scale factors is driven
by an additional positive effective cosmological constant, while the
other is driven by a negative one (unless $0<n<-1$, in which case both
effective cosmological constant terms can have the same sign).

It has been shown above that in all possible cases the two
FRW geometries effectively decouple from each other.  Their evolution
is then specified by a set of equations that are identical to
the Friedmann equation of General Relativity,
but with the possible addition of an effective cosmological constant term.  Further more, unless condition $(i)$ or
$(ii)$ is satisfied, we must also have $\hat{k}=k$ and $a \propto X^{1/4}Y^{3/4}$.

\subsection{Low acceleration limit}

The situation in the low acceleration limit is more complicated.  In
this case we have $\vert Q \vert \ll 1$ and
\be
\mathcal{M} \simeq \text{sign}(Q) \frac{4 \vert Q \vert^{3/4}}{3}+\mathcal{M}_1,
\ee
so that $\mathcal{M}_Q\simeq \vert Q \vert ^{-1/4}$.  Here $\mathcal{M}_1$ is
an unspecified constant.

Firstly, let us consider the situation with $k=\hat{k}=0$ and
$\mathcal{M}_1=0$ (we will return to the situation with non-zero
spatial curvature, and $\mathcal{M}_1 \neq 0$, below).  In this case
it can seen from Appendix C that we automatically have $Q=Q(\tau)$ and
$\sigma=\sigma(\tau)$, and that all dependence on $r$ vanishes from the
field equations.  Now, just as one can take $c \rightarrow \infty$ to
find the non-relativistic limit of General Relativity, here we take
$l\rightarrow 0$ to find the low acceleration limit of
the present theory.  Such a limit suppresses the
contribution of $G_{ab}$ and $\hat{G}_{ab}$ to the left hand side of
the field equations, relative to $S_{ab}$ and $\hat{S}_{ab}$,
so that (\ref{field1}) and (\ref{field2}) become
\be
S_{ab} \simeq 8 \pi G T_{ab}
\ee
and
\be
\hat{{\cal S}}_{ab} \simeq 8 \pi G \hat{T}_{ab}.
\ee
These equations are second-order, and
provide a system of constraint and evolution equations.
Assuming a perfect fluid form for both energy-momentum tensors we then
find the following matter dominated power-law solutions\footnote{By
 `matter dominated' we mean that the left-hand side of the field equations scales in the
same way as the right-hand side.  This is in contrast to `vacuum
dominated', in which the left-hand side scales independently of the
right-hand side, such that the influence of the matter fields on the
space-time dynamics is negligible.}:
\be
\label{prop}
a\propto \tau^p \qquad \text{and} \qquad X\propto Y\propto \tau^q, 
\ee
where $p$ and $q$ are given by
\bea
\label{p}
p &=& \frac{1-3 \hat{w}}{(1+2 w+8 n w-3 \hat{w}-8 n \hat{w}-6 w \hat{w})}\\
q &=& \frac{1-3 w}{(1+2 w+8 n w-3 \hat{w}-8 n \hat{w}-6 w
 \hat{w})},
\label{q}
\eea
and where $w$ and $\hat{w}$ are the equations of state of the two fluids, as
defined in Section \ref{FRWsec}.  These solutions are significantly
different from the corresponding solutions in General Relativity, and some
examples, for cosmologically interesting equations of state, are given
in Table \ref{tab1}.  In this table, and throughout, we use the time
coordinates $t$ and $\hat{t}$ to correspond to the proper time of comoving observers in each geometry,
so that $t = \int a d\tau$ and $\hat{t} = \int X d\tau$.

\begin{table}[htb]
\begin{center}
\begin{tabular}{|l|c|c|c|}
\hline
Fluids  & $\quad a(t)\quad$ &  $\quad Y(\hat{t})\quad$ & GR \\
\hline
dust \& dust & $t^{1/2}$ & $\hat{t}^{1/2}$  & $t^{2/3}$ \\
dust \& rad. & constant & $\hat{t}^{\frac{3}{(3-8n)}}$  & $t^{2/3}$ \&
$t^{1/2}$  \\
rad. \& dust & $t^{\frac{3}{8(1+n)}}$ & constant  & $t^{1/2}$ \&
$t^{2/3}$  \\
rad. \& rad. & $t^{\frac{3+8nC-3C}{8(1+n-C)}}$ & $\hat{t}^{C}$  & $t^{1/2}$\\
scalar \& scalar & $t^{1/4}$ & $\hat{t}^{1/4}$  & $t^{1/3}$ \\
$\Lambda$ \& $\Lambda$ & exponential & exponential  & exponential \\
\hline
\end{tabular}
\end{center}
\caption{
The functional form of matter dominated power-law solutions in the low
acceleration limit of Milgrom's bimetric theory and General Relativity
(GR).  The fluids considered are dust ($w=0$), radiation ($w=1/3$),
scalar fields ($w=1$) and $\Lambda$ ($w=-1$).  In the left hand
column, the first fluid is coupled to $g_{ab}$, and the second fluid
to $\hat{g}_{ab}$.  In the
radiation \& radiation case, the solutions are not specified uniquely,
and hence there is an arbitrary constant $C$ involved.
}  
\label{tab1}
\end{table}

It can be seen that the functional form of these solutions is, in
general, not only dependent on the fluid that is coupled to the metric
in question, but also to the fluid coupled to the other metric, as
well as the theory, via the parameter $n$.  This mutual dependence
should be expected as the two metrics are non-minimally coupled to
each other in the action.  The evolution of the two metrics
can also be asymmetric under the exchange of the two fluids.  This can be
seen from the second and third rows of Table \ref{tab1}.  Again, this
is expected, as $Q$ (as specified in Section \ref{spec}) is asymmetric
under the exchange of $g_{ab}$ and $\hat{g}_{ab}$.
Once these power-law solutions have been assumed, the field equations
then reduce to a set of algebraic relations between the $c_i$'s, $n$, $\alpha$, $\beta$, $l$, $\rho(t_0)$ and
$\hat{\rho}(t_0)$, as well as the constants of proportionality in
(\ref{prop}), where $t_0$ is the present age of the
Universe.

It is interesting to note that in some cases it is possible for one of the metrics to be
static.  This is exemplified in Table \ref{tab1} by the two radiation
\& dust solutions.  In fact, it can be seen from (\ref{p}) and
(\ref{q}) that if the matter coupled to the second metric is radiation,
then the first metric will be static (unless the denominator diverges,
as is the case with radiation \& radiation).  It is also
interesting to note that in the radiation \& radiation case the two
scale factors are not uniquely determined, and can only be specified up to
an arbitrary constant.  Clearly the behaviour of these solutions is
quite different to the usual general relativistic FRW solutions, which
are shown in the fourth column of Table \ref{tab1} for comparison.

A special case that should be noted is when $X$ equals $Y$, rather than simply
being proportional to it.  In this case the solutions that are
obtained are quite different to the general case, discussed above.
What happens is that the contributions from any terms that involve $Q$
vanish.  The contribution of $G_{ab}$ and $\hat{G}_{ab}$
can then no longer be considered to be suppressed, as the terms they were
suppressed with respect to no longer exist.  The system of equations
then reduces to exactly those considered previously, in the high
acceleration regime.  This is, in fact, the case that was considered
by Milgrom in \cite{Milgrom1}.

Now let us consider non-zero spatial curvature.  It can be seen from
Appendix C that $\sigma$ and $Q$ are, in general, functions of both
$\tau$ and $r$, when $k$ and $\hat{k}$ take arbitrary values.  Such a
dependence on $r$ is problematic, as the free functions in the
line-elements (\ref{ds}) and (\ref{ds2}) are all functions of $\tau$
only.  This is a generalization of the problem involving the $r$
dependence of $\sigma$ found in the high acceleration regime, and
although the dependence is more complicated, the solution is the same:
$\sigma$ and $Q$ reduce to functions of $\tau$ if $\hat{k}=k$.  It can then be seen that not only does all $r$
dependence drop out of $\sigma$, $Q$, $S_{ab}$ and $\hat{S}_{ab}$, but
that these functions also become independent of $k$ and $\hat{k}$ (up
to possible overall multiplicative factors of $(1-k r^2)$).  The only place that non-zero $k$ has any
effect in the field equations is therefore through its contribution to $G_{ab}$ and
$\hat{G}_{ab}$, where it appears as an additional constant.

Now, although the contribution of the $k$-independent terms in $G_{ab}$ are
suppressed by a power of $l$, and can therefore be safely ignored in
the low acceleration regime, the same cannot be said for $k$.  This
term is a constant, and does not have the potential to become
arbitrarily small at late-times, in contrast to the other terms in $G_{ab}$ and
$S_{ab}$.  We therefore cannot ignore it, as even if it is negligibly
small at early times, due to suppression by a power of $l$, it can
still become influential at late-times.  However, as the influence of
$k$ and $\hat{k}$ is limited to its appearance in the Einstein tensor,
when $\hat{k}=k$, its influence can be straightforwardly accounted
for:  It acts in exactly the same way as an additional fluid with
$w=-1/3$ coupled to $g_{ab}$, and a fluid with $\hat{w}=-1/3$ coupled
to $\hat{g}_{ab}$.  We can therefore calculate its influence at
late-times using equations (\ref{p}) and (\ref{q}), above.  We find
that the corresponding power-law solution has $a\propto X\propto Y \propto
\tau^3\propto t^{3/4}$, independent of $n$.  This should be contrasted
with the usual $a \propto t$ in General Relativity.

Next, let us consider the effect of non-zero $\mathcal{M}_1$. We
find that the inclusion of such a term requires $a\propto \tau^{-1}$, if
power-law solutions are to exist, which
correspond to exponential evolution in the proper time of comoving
observers, $t$.  The solutions that exist for $\hat{g}_{ab}$ then
depend on the matter content of the theory.  If we couple a fluid with
equation of state $w$ to $g_{ab}$ then we find the power-law solution
\be
X \propto Y \propto \tau^{-\frac{(3+4n+3w)}{4n}} \propto
\hat{t}^{1+\frac{4 n}{3 (1+w)}},
\ee
while if we couple a fluid with equation of state $\hat{w}$ to
$\hat{g}_{ab}$ we find
\be
X \propto Y \propto \tau^{-\frac{4(1+n)}{(1+4n-3\hat{w})}} \propto
\hat{t}^{\frac{4 (1+n)}{3 (1+\hat{w})}}.
\ee
Power-law solutions with fluids coupled to both metrics do not, in
general, exist, unless they result in $a\propto \tau^{-1}$ anyway (such as,
for example, two $\Lambda$ terms).

Finally, let us consider what happens on approach to a vacuum.  For
monotonically expanding geometries fluids with higher equations of
state give way to those with lower equations of state, such that dust
domination follows radiation domination, and so forth, in the usual
way.  If a fluid with equation of state $w=-1$ (or less) exists then
the total energy density will eventually reach a constant (or start to
increase), as all other fluids become negligible in comparison.  If no
such fluid exists, then the total energy density will decrease
monotonically forever.  If this happens then the right-hand side of
the field equations can eventually drop below the previously neglected
terms in the Einstein tensor.  For $p (1+3 w)>2$ this will mean that
the matter becomes subdominant, and the cosmological evolution will be
vacuum dominated (determined by the dynamics of the vacuum alone).  In this case the only
power-law solutions that exist go like $\tau^{-1}$, which corresponds to
exponential expansion.  For $p (1+3 w)<2$ the right-hand side of the
field equations will always be dominant over the neglected terms in
the Einstein tensor, and matter dominated power-law evolution, as described
above, can continue indefinitely into the future.  It should be noted
that if spatial curvature is non-zero, and acts like a fluid with
$w=-1/3$, then the condition $p (1+3 w)<2$ is always satisfied.  A
curvature dominated power-law solution, as found above, can therefore
always last for an indefinitely long period, provided there are no
fluids with equations of state $w<-1/3$.

\section{Viability as a Model of the Universe}
\label{viao}

We will now consider the viability of the cosmological solutions found
above as models of the observable Universe.  Our criterion for
viability will be whether or not it is possible to account for the
major probes of observational cosmology, which we take to be the
primordial synthesis of light elements, the position of the first
acoustic peak in the CMB angular power spectrum, the growth of
structure, and the late-time accelerating expansion of the Universe.
For later convenience, let us now define the density fraction of a
fluid $i$ to be
\be
\Omega_{i} \equiv \frac{8\pi G \rho_{i}}{3H_{0}^{2}},
\ee
where $\rho_{i}$ is the energy density of the fluid on a homogeneous
hypersurface of age $t_0$, and $H_{0}$ is the locally measured value
of the Hubble parameter (in whichever geometry the observers are taken
to be coupled to).

First consider the primordial synthesis of light elements.  This
process requires a cosmological evolution that is close to the
standard form \cite{ck}, $a\sim t^{1/2}$, and is highly sensitive to the ratio of photons to baryons,
$\eta$.  In particular, at the end of nucleosynthesis ($t \sim 200s$) it
can be shown that the observed abundances of light elements imply
$\eta = (5.5 \pm 0.5) \times 10^{-10}$ \cite{burles}.  If we now
combine this with a CMB photon temperature of $2.725\pm 0.001$K, and
assume that the expansion of the Universe has been adiabatic, then
we are left with the constraint 
\be
\label{baryons}
\Omega_{b} h^2 = 0.020 \pm 0.002,
\ee
to $95\%$ confidence, where $h\equiv H_0/(100 \text{km} \text{s}^{-1}
\text{Mpc}^{-1})$.  This is a tight constraint on the matter content of
the Universe, and one that we expect to be applicable to any viable model.

Next let us consider the positions of the acoustic peaks of the CMB.  In
general relativistic FRW cosmology their
angular extent on the sky, $\theta$, is essentially determined by two factors: The
acoustic horizon at last scattering, $R_{ls}$, and the angular diameter distance
to the last scattering surface, $d^*_A$, such that
\be
\label{l}
\theta \sim \pi \frac{d^*_A}{R_{ls}}.
\ee
Here the situation may be somewhat different.  The formation of linear
structure up to last scattering can be modified, as the
weak-field limit can now be altered from its usual Newtonian
form. This may affect the scale of correlations on the last scattering
surface, but the projection of that surface onto our sky will be
unaffected by this, and will therefore be sensitive to the intervening geometry of
space-time in the usual way.  We should therefore expect an equation
similar in form to (\ref{l}), but with $R_{ls}$ now replaced by
$\tilde{R}_{ls}$, referring to the new quantity that sets the scale of
correlations on the last scattering surface (no longer necessarily
the acoustic horizon).

A reliable calculation of $\tilde{R}_{ls}$ will require a detailed
analysis of linear perturbation theory, and pre-recombination physics,
within the frame-work of the current theory.  For our current
purposes, we will simply assume $\tilde{R}_{ls}$ has been calculated.
Using the well known result that spatial curvature produces a shift in
the angular scale of the first acoustic peak according to
$(1-\Omega_k)^{-9/20}$,
and given the compatibility of the observed CMB with a spatially flat
general relativistic FRW cosmology, we can then speculate that, in the
context of the present theory, the position of the first acoustic peak
is likely to imply an effective spatial curvature of
\be
\label{curvature}
\Omega_k \simeq 1- \left( \frac{\tilde{R}_{ls}}{R_{ls}} \right)^{20/9}.
\ee
This can be seen to reduce to $\Omega_k \simeq 0$ when
$\tilde{R}_{ls}\simeq R$.

Beyond the sensitivity of CMB observations to the
curvature of the Universe, the growth of structure also depends on
the evolution of the Universe.  This is true for both the linear
structure that we observe at last scattering, as well as the
subsequent growth of non-linear structures such as galaxies, and clusters of
galaxies.  In most cosmological models, it is usually required that
one should have a period of cosmological expansion where $a\sim
t^{2/3}$ in order for structure to form.  Here the situation may be
somewhat different, as, once again, the weak-field limit is modified
from its usual Newtonian form.  Never the less, we will consider it
preferable if the cosmological solutions we have found can be shown to
exhibit a period where $a$ or $X \sim t^{2/3}$.

Finally, let us consider supernova observations.  The Hubble diagrams
constructed from these events allow us to determine the
late-time evolution of the Universe.  In the standard cosmological
model, with the assumption of spatial flatness, they provide strong,
direct evidence for a cosmological constant, or vacuum energy, with
$\Omega_{\Lambda}\sim 0.7$ \cite{perl, riess}.  If one permits
non-zero spatial curvature, as will be required in these models if
$\tilde{R}_{ls}\neq R_{ls}$, then the resulting bounds on
$\Omega_{\Lambda}$ change.  However, it can still be
shown that there exists good evidence for a non-zero cosmological
constant, independent of the value of $\Omega_k$.  Using bounds derived
from the $414$ supernovae of the ``Union'' dataset one can make the
approximate statement that \cite{kowalski}
\be
\label{lambda}
0.3 \lesssim \Omega_{\Lambda} \lesssim 1.4, 
\ee
to $3 \sigma$, independent of the amount of spatial curvature\footnote{The precise
bounds achievable depend on exactly
which data one chooses to include, and how it is treated.}.  Any
viable model of the Universe should therefore include a late-time period
of accelerating expansion.

Having discussed the constraints available from observational
cosmology, let us now consider their applicability to the solutions
found in Section \ref{solutions}.

%

\subsection{High acceleration limit}

It was shown above that when $\vert Q \vert \gg 1$ the gravitational field
equations reduce to their usual general relativistic form, with a
cosmological constant term.  For a cosmological solution with dust and
radiation we will therefore have periods of cosmological evolution
with $a\sim t^{1/2}$ and $t^{2/3}$, when these two fluids
dominate the cosmological dynamics, respectively.  This is in good
keeping with the requirements of primordial nucleosynthesis and
structure formation.

If we apply the primordial nucleosynthesis bound (\ref{baryons}) on $\Omega_B$,
and take this to make up the entire contribution to
the dust content of the Universe, then the supernovae results \cite{kowalski} imply a
low value of $\Omega_{\Lambda} \sim 1/3$, that is
only compatible with observations at around the $3\sigma$ level.  What
is more, without any other matter content this would imply $\Omega_k
\sim 2/3$, which would only be compatible with the position of the
first acoustic peak of the CMB angular power spectrum if
$\tilde{R}_{ls} \sim 0.6 R_{ls}$.  Such a large change of scale seems
unlikely from previous studies using similar theories
\cite{pedro1,pedro2}.  We therefore find that some additional
non-baryonic dust-like degrees of freedom are necessary, either in the form of
cosmological dark matter, or from some other sector of the theory.

An effective cosmological constant term, $\mathcal{M}_0$, is also
present in (\ref{F1}) and (\ref{F2}).  This term has the attractive
feature of being able to reproduce the value of the
cosmological constant observed in supernova data with $O(1)$ values of $\mathcal{M}_0$ and $n$, for
astrophysically interesting values of $l$.  It also provides a lower
limit for $\vert Q \vert$, as $Q \rightarrow$constant in the limit
that $\mathcal{M}_0$ dominates.  This constant is zero in the
special case $X=Y$, and, as already discussed, in this case the
dynamics in the low acceleration limit become identical to those of
the high acceleration regime.  More generally we find that in the
limit $a\rightarrow a_0 \tau^{-1}$, $X\rightarrow X_0 \tau^{-1}$ and
$Y\rightarrow Y_0 \tau^{-1}$ we have
\be
Q \rightarrow - \frac{3 (3 c_3+c_5) l^2 (X_0^2-Y_0^2)^2}{a_0^2 X_0^4}.
\ee
If this value of $\vert Q \vert$ is $\gg 1$,
and the Universe starts off in the high acceleration regime, then the
whole of cosmological history could have taken place there.  In this case,
as discussed above, some non-baryonic cosmological dark matter would
be required. Alternatively, the decelerating phase of cosmological expansion could push the value
of $Q$ into the low acceleration regime at some point, depending on
parameters.  One may then look to the low or intermediate
acceleration regimes for the additional dust-like degrees of freedom.

As a further possibility, we note that one could consider including extra
interaction terms in the action that would behave like an extra dust-like
contribution during the matter dominated era, while staying in the
high acceleration regime. Indeed, such a contribution was found by the
authors of \cite{p3} when they considered a bimetric theory with the
interaction term
\begin{equation}
I_{int} \sim \int \sqrt{-\hat{g}}(\hat{g}^{-1})^{ab}g_{ab} d^{4}x.
\end{equation}
One would expect that the addition of such a term in the present
theory should have a similar effect in the $\vert Q \vert \gg 1$ regime,
but it is not immediately clear what effects this would have on
the weak-field limit.

\subsection{Low acceleration limit}

As discussed above, it seems entirely plausible that even if the
Universe starts off in the high acceleration regime (with $\vert Q
\vert \gg 1$), it could end up in the low acceleration regime (with
$\vert Q \vert \ll 1$) after a suitable period of decelerating expansion.
However, because $Q$ depends on the $c_{i}$'s, as well as
Hubble-parameter like terms and $l$, the values of
$H_{0}$ and $l$ do not necessarily tell us anything about the 
value of $Q$ today. At an extreme, if the $c_{i}$'s are
chosen appropriately, the Universe could be in the low acceleration limit even
before recombination, independent of $H_0$ and $l$. Thus, one could
also envisage a situation where the Universe is in the low acceleration
regime throughout its entire history.

Now let us consider what the probes of observational cosmology,
discussed above, can tell us about the possibility of cosmological
expansion in the low acceleration limit.  It can be seen from Table
\ref{tab1} that in this regime a period of cosmological expansion of the form $a\sim
t^{1/2}$ can be achieved as a result of dust-like
fluids being coupled to each of the two metrics.  We therefore have the
very unfamiliar situation of primordial nucleosynthesis 
being able to occur during a period of dust domination.  However, it can also
be seen that $a \sim t^{1/2}$ can be achieved for a variety of
other fluids coupled to the two metrics, including radiation, if we
allow $Y \sim \hat{t}^{\frac{(1+4 n)}{(1+8 n)}}$.

Now consider structure formation.  It can be seen that a period of
expansion of the form $a \sim t^{2/3}$ can be achieved if two
fluids with equation of state $w=-1/4$ are coupled to each of the two
metrics.  Having to invent two exotic fluids in this way is somewhat
distasteful, but again we have some freedom.  If we are prepared to
consider a second fluid with $\hat{w}=(3+16 n)^{-1}$,  then we can have $a
\sim t^{2/3}$ being produced with a fluid of dust coupled to the
first metric.  For a theory with $n=-1/8$ this corresponds to dust and
a scalar field.  Various other situations can be read off from
Equations (\ref{p}) and (\ref{q}).  One particularly interesting
example is for a theory with $n=-1/16$. In this case spatial curvature dominating the first metric, and
dust coupled to the second metric, also gives an evolution
$a \sim t^{2/3}$.  Here, then, the dust coupled to the second
metric acts, in a way, as if it were cosmological dark matter coupled to
the first, which is itself an empty, open universe.  Clearly there is
considerable scope for new and interesting behaviour.

\subsection{The intermediate regime}

One may also ask whether a dust-like contribution could arise from the
intermediate regime, where $\vert Q \vert \sim 1$. As only
the asymptotic limits of this function are defined, there is considerable
flexibility in terms of the function's transitional behaviour. For
example, if we consider a transitional regime where $\mathcal{M} \sim
\vert Q \vert^s$ then an analysis similar to that used in the low
acceleration regime finds solutions $a\propto \tau^p$ and $X\propto
Y\propto \tau^q$ where
\bea
\nonumber
p &=& \frac{2 s (1-3 \hat{w})}{(3-2s+3 w+12 n w-9 \hat{w}-12 n
 \hat{w}+6 s \hat{w}-9 w \hat{w})}\\
q &=& \frac{2 s (1-3 w)}{(3-2s+3 w+12 n w-9 \hat{w}-12 n \hat{w}+6 s
 \hat{w} -9 w\hat{w})}.
\nonumber
\eea
These equations reduce to (\ref{p}) and (\ref{q}) when $s=3/4$.  It
can be seen from the above that, among very many other possibilities,
we can arrange to have $a \sim t^{2/3}$ and $Y\sim
\hat{t}^{2/3}$ when $w=\hat{w}=s-1$.  For a theory in which $s=2/3$,
we can therefore have the some of the effects of cosmological dark
matter when both metrics are effectively empty, with negative spatial
curvature.  This is intended as an example only.  Without any well defined functional form of
$\mathcal{M}(Q)$ in the intermediate regime one clearly has
considerable freedom to achieve whatever evolution is desired.

One final possibility arises if the entirety of cosmological history
corresponds to values of $Q$ with a different sign to that which is
appropriate for the weak-field limit.  Here we have considered the
functional form of ${\cal M}(Q)$ to be set by the magnitude of $Q$
only.  If one allowed its form to be different for positive or
negative $Q$, then the possibility could arise that the
form of ${\cal M}(Q)$ that is appropriate for cosmology is not fixed at all by
considerations of the weak-field limit.  Cosmological
solutions could then be considered entirely independently from other
phenomenology.  In this case one may look to the discussion of
$\vert Q \vert \sim 1$, above, for what the
cosmological dynamics could look like for simple power-law forms of
$\mathcal{M}(Q)$.

\section{Discussion}
\label{disc}

We have studied the FRW solutions of Milgrom's class of bimetric
theories of gravity.  These theories have different behaviours
depending on the value of the scalar quantity $\vert Q \vert$, which
is formed from the two metrics and various parameters of the theory.
The two regimes that result are referred to as the `high' and `low acceleration' limits.

We find that in the high acceleration
limit the cosmological dynamics of the two metrics essentially
decouple from each another, and evolve in a similar fashion to the FRW
solutions of General Relativity.  In this regime, the theory also
potentially provides an explanation of the cosmological constant
problem, as it is possible to include a constant term in the Friedmann
equations that is constructed from a factor of order unity divided by
the square of the intrinsic length scale of the
theory, $l$.  This results in the correct order of magnitude for the cosmological
constant measured in our observable Universe.  Solutions that stay in
the high acceleration regime for their entire history, however, appear to be in
conflict with observations, unless non-baryonic dark matter fields, or
extra interaction terms in the action, are also included.

In the low acceleration limit we find that simple power-law, matter
dominated solutions also exist.
There is considerable freedom in the form of these solutions,
depending on the parameters of the theory, as well as the matter
fields coupled to each of the two metrics.  The solutions in this
regime are, in general, quite different to their general relativistic
counterparts, and display some interesting possibilities.  For
example, in some theories it is possible to have a metric that
evolves like a dust dominated universe does in General Relativity,
while itself being empty and spatially open.  This allows for the
possibility of accounting for some of the effects of cosmological dark
matter via the fields coupled to the second metric.
\vspace{-20pt}
\section*{Acknowledgements}

We are grateful to M. Ba\~{n}ados, P. G. Ferreira and T. Jacobson for
helpful comments and discussions.
TC acknowledges the support of Jesus College, Oxford and the BIPAC.
Research at Perimeter Institute is supported by the Government of Canada through Industry Canada and
by the Province of Ontario through the Ministry of Research and Innovation.

\section*{Appendix A: \quad Derivation of the Field Equations}

To derive the field equations, let us consider the variations of $\mathcal{M}$, $\sigma$,
and $\sqrt{-g}$ with respect to $g^{ab}$ separately.  Firstly, varying
$\mathcal{M}$ gives
\begin{eqnarray}
\sigma \delta \mathcal{M} &=& \sigma l^2 \mathcal{M}_{Q}\frac{\delta
 Q_{ad}^{\phantom{ad}bcef}}{\delta
 g^{rs}}C^{a}_{\phantom{a}bc}C^{d}_{\phantom{d}ef}\delta g^{rs}
\label{dm1} \\
\nonumber &&  +2 \sigma l^2 \mathcal{M}_{Q}J_{d}^{\phantom{d}ef}\delta \Gamma^{d}_{\phantom{d}ef}
\end{eqnarray}
where $\mathcal{M}_{Q}\equiv d\mathcal{M}/dQ$, and $J_{d}^{\phantom{d}ef} \equiv Q_{ad}^{\phantom{ad}bcef}
C^{a}_{\phantom{a}bc}$.  The second term in the equation above can
then be written
\begin{eqnarray}
\nonumber
&& 2 \sigma l^2 \mathcal{M}_{Q}J_{d}^{\phantom{d}ef}\delta
\Gamma^{d}_{\phantom{d}ef} \\&=& \sigma l^2
\mathcal{M}_{Q}J_{d}^{\phantom{d}ef}g^{d z}(\delta g_{z f;e}
           +\delta g_{ez;f} -\delta g_{ef;z})
	    \nonumber\\
&\dot{=}& l^2 \left[\sigma
	    \mathcal{M}_{Q}\left( J_{(ab)}^{\phantom{(ab)}e}+J_{(a\phantom{e}b)}^{\phantom{(a}e}
	    -J^{e}_{\phantom{e}ab}\right)\right]_{;e} \delta g^{ab}, \label{dm2}
\end{eqnarray}
where $\dot{=}$ means equal up to total divergences, and where we have used
the usual result $\delta g_{ef} = - g_{ea}g_{fb}\delta g^{ab}$.  The
quantities $J_{ab}^{\phantom{ab}e}$ and $J^{e}_{\phantom{e}ab}$, in
the equation above, have had their indices lowered with $g_{ab}$.

We also have
\be
\label{dg}
\sigma \mathcal{M} \frac{\delta (\sqrt{-g})}{\sqrt{-g}} =-\frac{\sigma}{2} \mathcal{M} g_{ab}\delta g^{ab},
\ee
and
\be
\mathcal{M} \delta \sigma = \frac{\delta{\sigma}}{\delta g^{ab}}\mathcal{M} \delta g^{ab}.
\ee
If $\sigma=\kappa^{2 n}$ then $\delta \sigma=-\frac{n}{2}\sigma g_{ab}\delta g^{ab}$,
and the last equation can then be written
\be
\label{dsigma}
\mathcal{M} \delta \sigma = -\frac{n}{2}\sigma g_{ab} \mathcal{M} \delta g^{ab}.
\ee
Summing (\ref{dm1}), (\ref{dg}) and (\ref{dsigma}) gives
the field equations (\ref{field1}) and (\ref{field2}) shown in Section \ref{field}.

\section*{Appendix B: \quad The Constraint Equations}

Given the complicated form of the full second-order field equations,
it is useful to look for the simpler, first-order constraint equation.
To find this, consider general variations of the action of the form
\begin{eqnarray}
\label{varo}
\delta I &=& \int d^{4}x\sqrt{-g}(E_{ab}\delta g^{ab} +\hat{E}_{ab}\delta \hat{g}^{ab}),
\end{eqnarray}
where 
\be
E_{ab} = \beta G_{ab}-S_{ab}-\frac{1}{2}T_{ab} = 0
\ee
and
\be
\hat{E}_{ab} = \sqrt{\frac{\hat{g}}{g}} (\alpha \hat{G}_{ab}-\hat{S}_{ab}-\frac{1}{2}\hat{T}_{ab})=0.
\ee
If we assume that $T_{ab}$ and $\hat{T}_{ab}$ contain at most
first-order derivatives then we do not need to consider any compensatory contributions from their evolution equations
in order to find the desired constraint equations.

Now consider the variation of $g^{ab}$ and
$\hat{g}^{ab}$ due to diffeomorphisms generated by the
`infinitesimal' vector field $\xi^{a}$. 
%
%
%
%
Firstly, let us consider the variation of $\delta g^{ab}$ due to
$\xi^{a}$, which gives
\begin{eqnarray}
\delta I_{\xi,\delta g} &=& -\int E_{ab}(g^{ac}\nabla_{c}\xi^{b}+g^{cb}\nabla_{c}\xi^{a}).
\end{eqnarray}
We can integrate this by parts to yield a boundary term, that we
neglect, the Bianchi identity, $\nabla_{a}G^{ab}=0$, and
\begin{eqnarray}
\delta I_{\xi,\delta g} &=& 2\int \nabla_{c}(E^{ca}) \xi_{a}.
\end{eqnarray}
Now consider variations of $\hat{g}^{ab}$, such that
\begin{eqnarray}
\delta I_{\xi,\delta \hat{g}} &=& \int \hat{E}_{ab}(\xi^{c}\nabla_{c}\hat{g}^{ab}-\hat{g}^{ac}\nabla_{c}\xi^{b}-\hat{g}^{cb}\nabla_{c}\xi^{a})\\
\nonumber &=& \int
(\xi_{c}(\nabla^{c}\hat{g}^{ab})\hat{E}_{ab}+\xi_{a}\nabla_{c}(\hat{E}_{b}^{\phantom{b}a}
\hat{g}^{bc}+\hat{E}^{a}_{\phantom{b}b}\hat{g}^{cb})).
\end{eqnarray}
Collecting terms then gives
\begin{eqnarray}
\delta I_{\xi} &=& I_{\xi,\delta g}+I_{\xi,\delta\hat{g}} \\
\nonumber &=& \int \xi_{a}(
(\nabla^{a}\hat{g}^{cb})\hat{E}_{cb}+\nabla_{c}(
2E^{ca}+\hat{E}_{b}^{\phantom{b}a}
\hat{g}^{bc}+\hat{E}^{a}_{\phantom{b}b}\hat{g}^{cb})).
\end{eqnarray}
Finally, let us define the tensor  ${\cal J}^{ca} \equiv
2E^{ca}+\hat{E}_{b}^{\phantom{b}a}
\hat{g}^{bc}+\hat{E}^{a}_{\phantom{b}b}\hat{g}^{cb}$. If ${\cal J}^{0a}$ had
second-order derivatives, then $\nabla_{c}{\cal J}^{ca}$ would
contain third-order derivatives. However, we know from
the vanishing of the variation of the action under diffeomorphisms, $\delta I_{\xi}=0$,
that $\nabla_{c}{\cal J}^{ca} =- (\nabla^{a}\hat{g}^{cb})\hat{E}_{cb} $.
Now, due to the structure of the theory we can see that $\hat{E}_{cb}$ contains at most second-order
derivatives, so the right-hand side of this equation can contain at most
second-order derivatives. We can therefore conclude that ${\cal J}^{0a}$ contains
up to first-order derivatives only. 

For a given foliation of hypersurfaces with normal $n^{a}$, and metric
$h_{ab}=g_{ab}+n_{a}n_{b}$, we then find that the constraint equations
are given by
\be 
{\cal J}_{ab}n^{a}h^{b}_{\phantom{b}c} = 0.
\ee
and
\be
{\cal J}_{ab}n^{a}n^{b} = 0
\ee
The first of these is trivially satisfied by FRW geometry, where the
hypersurfaces are taken to be surfaces of constant $t$, while the
second gives the constraint equations displayed in Appendix C, and
Section \ref{FRWsec}.

\section*{Appendix C: \quad The Field Equations with FRW Geometry}

Here we will write explicit expressions for some of the quantities that appear in the field
equations (\ref{field1}) and (\ref{field2}), for the FRW geometries
(\ref{ds}) and (\ref{ds2}).  Firstly, we can immediately write
\be
\label{sigapp}
\sigma = \left ( \frac{\sqrt{1-\hat{k} r^2}}{\sqrt{1-k r^2}}
\frac{a^4}{XY^3} \right)^n.
\ee
We can also write a relatively simple expression for $Q$.  To do this
we first define five new quantities via
\be
\label{Qapp}
Q=c_1 Q^{(1)}+c_2 Q^{(2)}+c_3 Q^{(3)}+c_4 Q^{(4)}+c_5 Q^{(5)},
\ee
such that $Q^{(1)}
= l^2 \delta^f_{\phantom{f} a}
 \delta^d_{\phantom{d} b} g^{c e} C^a_{\phantom{a} cd}
 C^b_{\phantom{b} ef}$,
with the other four defined {\it mutatis mutandis}.  We can then write
\begin{widetext}
\vspace{-10pt}
\begin{eqnarray*}
Q^{(1)} &=& \frac{l^2}{a^2} \left[ 2 \frac{\dot{a}^2}{a^2}+ 2
 \frac{\dot{a}}{a}\frac{\dot{X}}{X} - \frac{\dot{X}^2}{X^2}-3
 \frac{\dot{Y}^2}{Y^2} +\frac{(k-\hat{k})^2 r^2}{(1-k r^2) (1-\hat{k}
   r^2)^2} + \frac{2 ((1-k r^2)+2 (1-\hat{k}r^2))}{(1-\hat{k}r^2)}
 \frac{Y \dot{Y}}{X^2} \left( \frac{\dot{Y}}{Y}-
 \frac{\dot{a}}{a}\right) \right]\\
Q^{(2)} &=& \frac{l^2}{a^2} \Bigg[ 8 \frac{\dot{a}^2}{a^2}+2
 \frac{\dot{a}}{a} \frac{\dot{X}}{X} -6 \frac{\dot{a}}{a}
 \frac{\dot{Y}}{Y} - \frac{\dot{X}^2}{X^2} -3
 \frac{\dot{X}}{X}\frac{\dot{Y}}{Y} + \frac{(k-\hat{k})^2 r^2 (3-2
 \hat{k} r^2)}{(1-k r^2) (1-\hat{k}r^2)^2}\\&\;&\qquad\qquad \qquad\;\;\qquad \qquad \qquad \qquad\qquad\qquad \qquad+ \frac{((1-k r^2)+2
 (1-\hat{k}r^2))}{(1-\hat{k} r^2)} \frac{Y \dot{Y}}{X^2} \left(
 \frac{\dot{X}}{X}+3\frac{\dot{Y}}{Y} -4 \frac{\dot{a}}{a}
 \right) \Bigg]
\\
Q^{(3)} &=& \frac{l^2}{a^2} \Bigg[ - 4 \frac{\dot{a}^2}{a^2}-4
 \frac{\dot{a}}{a}\frac{\dot{X}}{X}-\frac{\dot{X}^2}{X^2} + \frac{(k-\hat{k})^2 r^2 (3-2
 \hat{k} r^2)^2}{(1-k r^2) (1-\hat{k}r^2)^2}\\&\;&\qquad \;\;\;\;\qquad
 \qquad \qquad+ \frac{((1-k r^2)+2
 (1-\hat{k}r^2))}{(1-\hat{k} r^2)} \frac{Y \dot{Y}}{X^2} \left(
 4\frac{\dot{a}}{a}+2 \frac{\dot{X}}{X}- \frac{((1-k r^2)+2
 (1-\hat{k}r^2))}{(1-\hat{k} r^2)} \frac{Y \dot{Y}}{X^2} \right) \Bigg]
\\
Q^{(4)} &=& \frac{l^2}{a^2} \Bigg[ -16 \frac{\dot{a}^2}{a^2}+8
 \frac{\dot{a}}{a}\frac{\dot{X}}{X}+24
 \frac{\dot{a}}{a}\frac{\dot{Y}}{Y}-\frac{\dot{X}^2}{X^2}-6
 \frac{\dot{X}}{X}\frac{\dot{Y}}{Y}-9 \frac{\dot{Y}^2}{Y^2} +
 \frac{(k-\hat{k})^2 r^2}{(1-k r^2)(1-\hat{k} r^2)^2}\Bigg]
\\
Q^{(5)} &=& \frac{l^2}{a^2} \Bigg[ -10 \frac{\dot{a}^2}{a^2}+2
 \frac{\dot{a}}{a}\frac{\dot{X}}{X} +12
 \frac{\dot{a}}{a}\frac{\dot{Y}}{Y} -\frac{\dot{X}^2}{X^2}-6
 \frac{\dot{Y}^2}{Y^2} +\frac{(k-\hat{k})^2 r^2 (1+2
 (1-\hat{k}r^2)^2)}{(1-k r^2) (1-\hat{k}r^2)^2} \\ &\;&\qquad\; \qquad\qquad \qquad
 \qquad \qquad+ 2\frac{((1-k r^2)+2
 (1-\hat{k}r^2))}{(1-\hat{k} r^2)} \frac{Y \dot{Y}\dot{a}}{X^2a}  -\frac{((1-k r^2)^2+2
 (1-\hat{k}r^2)^2)}{(1-\hat{k} r^2)^2} \frac{Y^2 \dot{Y}^2}{X^2} \Bigg].
\end{eqnarray*}
When $\hat{k}=k$ (motivated in Section \ref{solutions}), the constraint equation can then be written for
arbitrary $c_1$, $c_2$, $c_3$, $c_4$ and $c_5$ in relatively simple form:
\begin{eqnarray*}
&& \beta \left(\frac{\dot{a}^2}{a^2}+k\right) + \alpha \frac{Y^3}{a^2 X}\left(\frac{\dot{Y}^2}{Y^2}+k\frac{X^2}{Y^2}\right)
+\frac{a^2 \mathcal{M} \sigma}{3 l^2}- \frac{8 \pi G
\rho a^2}{3}- \frac{8 \pi G \hat{\rho} X Y^3}{3 a^2}
\\&=& \frac{2\sigma \mathcal{M}_Q}{3} \Bigg[
2(c_1-2c_2-2c_3-8c_4-5c_5) \frac{\dot{a}^2}{a^2} +
(2c_1+5c_2-4c_3+8c_4+2c_5) \frac{\dot{a}}{a}\frac{\dot{X}}{X}\\&&\qquad\qquad\; -
   (c_1+c_2+c_3+c_4+c_5) \frac{\dot{X}^2}{X^2} + \frac{3((c_2+8c_4+4c_5)X^2
	+2(-c_1+2c_3+c_5)Y^2)}{X^2} \frac{\dot{a}}{a}
\frac{\dot{Y}}{Y} \\&&\qquad\qquad\;-\frac{3 ((c_2+2c_4)X^2-2 c_3Y^2)}{X^2}
\frac{\dot{X}}{X}\frac{\dot{Y}}{Y}-\frac{3((c_1+3c_4+2c_5)X^4-2
 c_1X^2Y^2+(3c_3+c_5)Y^4)}{X^4} \frac{\dot{Y}^2}{Y^2}\Bigg],
\end{eqnarray*}
where $\sigma$ and $Q$ are given by (\ref{sigapp}) and (\ref{Qapp}),
above.  The special case of the `concrete simple theory' of Milgrom is given
in Section \ref{FRWsec}.  Due to their length, we choose not to
display the evolution equations here, which make up the remaining part of the field equations.
\end{widetext}

\end{document}